\documentclass[reprint,amsmath,amssymb,aps]{revtex4-2}
\usepackage{CJK}
\usepackage{graphicx}
\usepackage{color}
\usepackage{amsmath}
\usepackage{amssymb}
\usepackage{amsthm}
\usepackage{extarrows}
\usepackage{tabularx}
\usepackage{dsfont}
\usepackage{booktabs}
\usepackage{float}
\usepackage[dvipdfm,colorlinks,urlcolor=blue,linkcolor=blue,anchorcolor=blue,citecolor=blue]{hyperref}
\usepackage{lineno}

\usepackage{appendix}

\begin{document}

\title{Heat Conduction in Momentum-Conserving Fluids: From quasi-2D to 3D systems}



\author{Rongxiang Luo$^{1}$}
\author{Jiaqi Wen$^1$}
\author{Juncheng Guo$^1$}

\affiliation{
\vspace{0.5em}
$^1$ Department of Physics, \href{https://ror.org/011xvna82}
{\color{blue}{Fuzhou University}}, Fuzhou 350108, Fujian, China
}

\date{\today }

\begin{abstract}
Using nonequilibrium and equilibrium molecular dynamics simulations, we investigate heat conduction in a momentum-conserving mesoscopic fluid modeled by multiparticle collision dynamics. Across quasi-two-dimensional (q-2D) to three-dimensional (3D) systems, we identify three distinct transport regimes: (i) a \emph{ballistic regime}, where thermal conductivity scales linearly with system size ($\kappa \sim L$) and the total heat current autocorrelation function $C(t)$ remains constant; (ii)~a \emph{kinetic regime}, characterized by size-independent $\kappa$ and exponentially decaying $C(t)$,  demonstrating that normal heat conduction dominated by kinetic effects is far more ubiquitous than previously observed in 1D systems; and (iii)~a \emph{hydrodynamic regime}, where the q-2D system exhibits logarithmically divergent conductivity ($ \kappa \sim \ln L $ ) with  $ C(t) \sim t^{-1} $ , while the 3D system displays finite  $ \kappa $  and  $ C(t) \sim t^{-3/2} $. Our results, observed in the hydrodynamic regime, quantitatively validate the scaling predictions for heat transport and reveal a clear dimensional crossover---from 2D-like anomalous transport to 3D Fourier behavior. These results lay a foundation for understanding thermal transport in q-2D to 3D systems and have practical implications for the design of micro- and nanoscale thermal devices.\par
\end{abstract}

\maketitle

\emph{Introduction}---A thorough understanding of heat conduction is essential for constructing microscopic pictures of macroscopic irreversible heat transfer and provides the theoretical foundation for thermal energy control and management~\cite{lepri2016,ZHANG20201,BENENTI20171,2012Li,RevModPhys.90.041002,RevModPhys.94.025002}. In 1D momentum-conserving systems, it is well known that the thermal conductivity $\kappa$ generally diverges with system size $L$ as $\kappa \sim L^{\alpha}$, signaling a breakdown of Fourier's law; however, numerical evidence has yet to conclusively settle the value of the exponent $\alpha$ predicted by various theories~\cite{Lepri1997,Dhar2001,Narayan2002,Casati2003,Li2003,Denisov2003,Cipriani2005,Delfini2007,Wang2011,Beijeren2012,Spohn2014,Hurtado2016,Lihb2010}. For high-dimensional momentum-conserving systems, it is predicted by different theoretical approaches~\cite{Narayan2002,Lepri1998} and by exactly solved stochastic models~\cite{Basile2006} that the heat current autocorrelation function $C(t)$ decays as a power law in time, $C(t) \sim t^{\beta}$, with $\beta = -1$ for the 2D systems and $\beta = -3/2$ for the 3D systems. These exponents have recently been quantitatively verified in a mesoscopic fluid model~\cite{e27050455}, and imply that $\kappa$ diverges logarithmically with system size $L$ in two dimensions but is finite in three dimensions. A further issue concerns dimensional crossover in heat conduction: substantial evidence~\cite{e27050455,Wang2010,Saito2010,Yang2006,Luo2020,Hisamoto2025,Dong2018} indicates a crossover from 3D or 2D to 1D heat conduction behaviors; yet, to our knowledge, conclusive evidence for a crossover from 3D to 2D conduction remains lacking.\par

Interestingly, normal heat conduction can emerge in 1D momentum-conserving systems, as demonstrated by a few models incorporating specific interactions or specific parameter regimes within a single model~\cite{PhysRevLett.84.2144,PhysRevLett.84.2381,Zhong2012,Zhong2013,Chen2016,Savin2014,Chen2014,Chenhb2025,zhao2018,Fu2026}. In particular, for 1D models with asymmetric interactions~\cite{Zhong2012,Zhong2013,Chen2016} or in the nearly integrable limit~\cite{Chen2014}, subsequent studies have shown that normal heat conduction in such systems is a finite-size effect dominated by kinetic processes~\cite{zhao2018,Fu2026}, which induces exponential decay of $C(t)$; anomalous hydrodynamic behavior is recovered for sufficiently large system size~\cite{Fu2026,Wang2013,Das2014,Miron2019,Lepri2020}. All real materials---whether one-, two-, or three-dimensional, or even exhibiting dimensional crossovers---are finite in size. A clear understanding of how kinetic and hydrodynamic regimes influence heat transport in more general, higher-dimensional systems is therefore essential. In this work, we address a fundamental question: How does heat conduction evolve as a system transitions from q-2D to fully 3D systems, and under what conditions does normal transport emerge?\par

Here, we aim to provide conclusive evidence resolving the above question. While molecular dynamics (MD) simulations of lattice models offer the most natural microscopic description~\cite{Wang2010,Saito2010}, they suffer from a key limitation in large 3D systems: thermal fluctuations result in prohibitively high computational costs, thereby preventing quantitative convergence of transport coefficients. To overcome this limitation, we employ multiparticle collision (MPC) dynamics~\cite{1999Malevanets}---a mesoscale approach in which point particles undergo stochastic collisions conserving total momentum and kinetic energy. A key advantage of this method is its ability to efficiently capture both thermal fluctuations and hydrodynamic interactions, with significant computational efficiency~\cite{2009Multi,PhysRevE.74.031402}. The MPC method has been successfully applied to study fundamental problems in statistical physics~\cite{Belushkin2011,Benenti2014,Luo2018,1Luo2020,Luo2025}, and in particular, the effect of external sources.\par

Using the MPC method, we uncover a rich phase diagram of heat conduction governed by the interplay between interaction strength and effective dimensionality. Our combined nonequilibrium and equilibrium simulations reveal three universal regimes: ballistic, kinetic, and hydrodynamic. Remarkably, we find that normal heat conduction, characteristic of the kinetic regime, persists from q-2D to 3D systems under weak interactions, substantially extending the domain of validity of Fourier's law. Conversely, strong interactions trigger a dimensional crossover: while 3D systems exhibit size-independent $\kappa$, q-2D systems exhibit the hallmark 2D-like anomaly $\kappa \sim \ln L$. These findings establish a foundation for understanding thermal transport across q-2D  to 3D systems and may have practical implications for the design of micro- and nanoscale thermal devices.\par
\par

\emph{The Model}---The system we consider comprises $N$ identical point particles of mass $m$, confined to a cuboid domain of length $L$, width $W$, and height $H$ in Cartesian coordinates $(x,y,z)$. The system is coupled to two stochastic thermal walls of area $WH$, located at the boundaries $x=0$ and $x=L$; each wall reflects incident particles with a newly sampled velocity $\bm v = (v_x, v_y, v_z)$ drawn from the following distributions~\cite{1978Transport,PhysRevE.57.R17}:
\begin{equation}\label{Eqfv}
\begin{aligned}
  f_{\alpha}(v_x) &= \frac{m |v_x|}{k_{\mathrm{B}} T_\alpha} \exp\!\left(-\frac{m v_x^2}{2k_{\mathrm{B}} T_\alpha}\right), \\
  f_{\alpha}(v_{y,z}) &= \sqrt{\frac{m}{2\pi k_{\mathrm{B}} T_\alpha}} \exp\!\left(-\frac{m v_{y,z}^2}{2k_{\mathrm{B}} T_\alpha}\right),
\end{aligned}
\end{equation}
where $\alpha = 0,L$ labels the heat bath and $k_{\mathrm{B}}$ is Boltzmann's constant. The sampling enforces the physical constraint that $v_x > 0$ for reflections at $x=0$ and $v_x < 0$ for reflections at $x=L$; $v_{y,z}$ remains unrestricted. Periodic boundary conditions are imposed in the $y$- and $z$-directions.\par

The bulk dynamics is described by the MPC method~\cite{1999Malevanets,2009Multi}. In the MPC method, the system evolves in discrete time steps, consisting of non-interacting propagation over a time interval $\tau$, followed by instantaneous collision events. During propagation, each particle retains its velocity $\bm v_i$, and its position is updated as
\begin{equation}\label{ri}
 \bm r_i\to \bm r_i + \bm v_i\tau.
\end{equation}
For collisions, the system volume is partitioned into cubic cells of linear size $a$. Within each cell, the velocities of all particles are rotated about a randomly chosen axis relative to the cell's center-of-mass velocity $\bm V_{\mathrm{CM}}$ by an angle $\theta$ or $-\theta$, each selected with equal probability. The velocity of particle $i$ is thus updated as
\begin{equation}\label{vi}
 \bm v_i \to \bm V_{\mathrm{CM}} + \hat{\mathcal{R}}^{\pm\theta} \bigl( \bm v_i - \bm V_{\mathrm{CM}} \bigr),
\end{equation}
where $\hat{\mathcal{R}}^{\theta}$ denotes the rotation operator by angle $\theta$. To ensure Galilean invariance of the stochastic rotation dynamics, the collision cells are randomly shifted before each collision step~\cite{PhysRevE.63.020201}. This collision rule conserves both the total momentum and total kinetic energy of the particles within each cell. The parameter $\tau$---the time interval between successive collisions---controls the effective interaction strength: smaller $\tau$ implies more frequent collisions and hence stronger inter-particle coupling. In the noninteracting case, $\tau = \infty$. Consequently, increasing $\tau$ suppresses collisional interactions and thereby tunes the transport properties of the system.\par

In the nonequilibrium setup, we set $T_\alpha$ to be slightly offset from the nominal temperature $T$, i.e., $T_{0,L} = T \pm \Delta T/2$, to investigate the dependence of the thermal current $j$ on the system length $L$. In our simulations, each particle is initially assigned a random position uniformly distributed in the volume and a random velocity drawn from the Maxwell--Boltzmann distribution at temperature $T$. After the system reaches the steady state, we compute the thermal current $j$ as the net energy flux per unit area and unit time transferred from the thermal wall to the system, and substitute $j$ into Fourier's law,
$\kappa = \frac{j L}{T_0 - T_L}$,
to obtain $\kappa$. We thus examine the dependence of $\kappa$ on $L$ to determine whether heat conduction is anomalous or normal. All data points shown are from numerical simulations, with statistical uncertainties (relative standard errors) below $1\%$, smaller than the symbol size.\par

\begin{figure}[t]
  \centering
  \includegraphics[width=1\columnwidth]{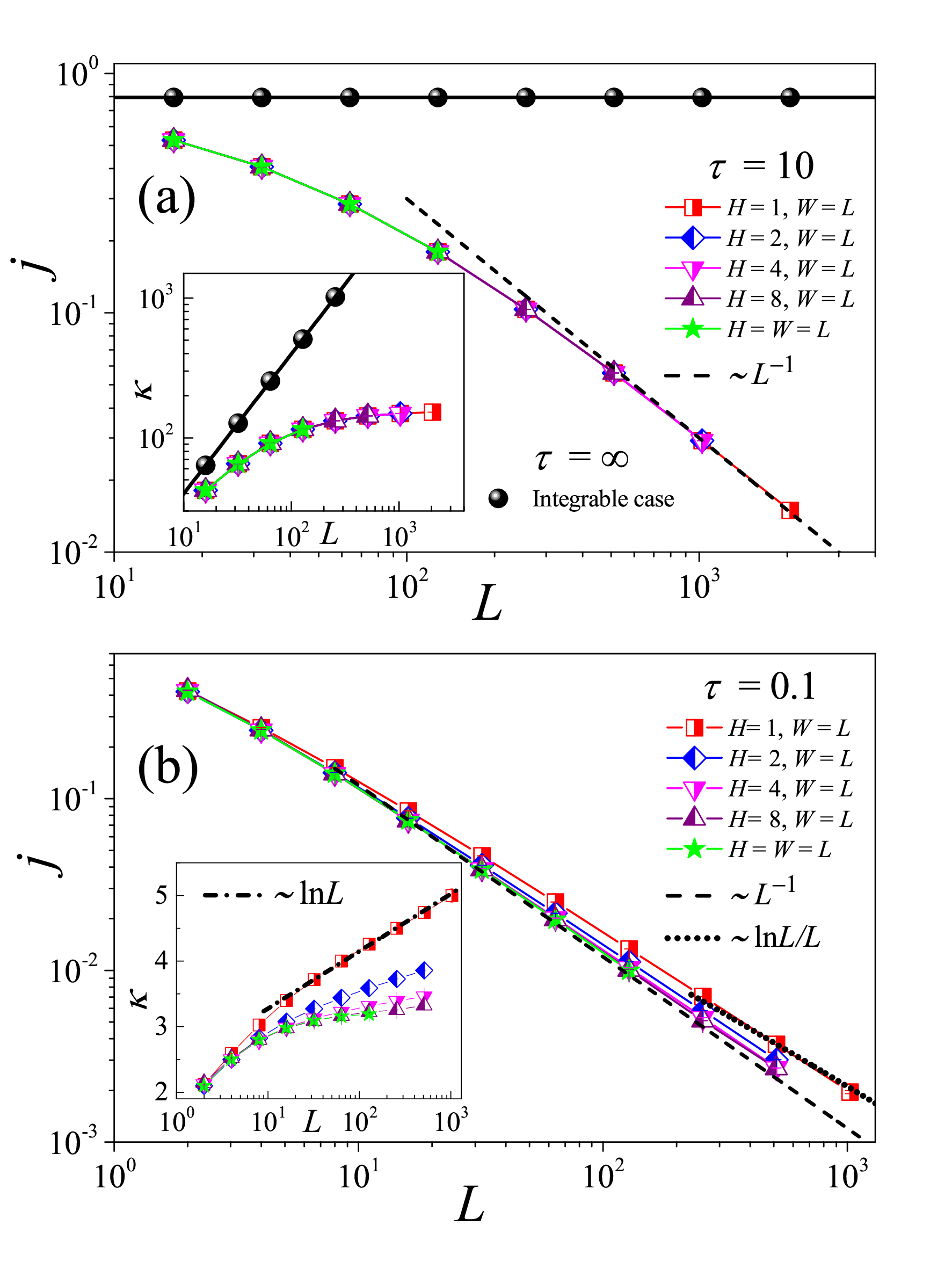}
  \caption{Dependence of the thermal current $j$ and thermal conductivity $\kappa$ on system size $L$ for three values of $\tau$. The black solid lines in (a) and its inset correspond to the analytical results in Eqs.~(\ref{Eq4}) and (\ref{Eq5}), obtained in the noninteracting (integrable) limit. Dashed: $j \sim L^{-1}$; dotted: $j \sim \ln L/L$; dot-dashed: $\kappa \sim \ln L$. All simulations are performed in reduced units, with $m = k_{\mathrm{B}} = T = a = 1$, $\Delta T = 0.2$, $\theta = \pi/2$, and $\rho = 6$.}
  \label{fig1}
\end{figure}

\emph{Non-equilibrium results}---We first provide analytical and numerical evidence that, in the ballistic regime, where the system is integrable, the thermal current is constant and the thermal conductivity scales linearly with system size. In the noninteracting case ($\tau = \infty$), particles traverse the system without collisions, retaining their initial velocities---rendering the system integrable. Following the analysis of Ref.~\cite{PhysRevE.99.032138}, we obtain an analytical expression for the thermal current~\cite{PhysRevResearch.4.043184}:
\begin{equation}\label{Eq4}
j = \sqrt{\frac{8\rho^{2} k_{\mathrm{B}}^{3}}{\pi m}} \, \sqrt{T_0 T_L} \left( \sqrt{T_0} - \sqrt{T_L} \right),
\end{equation}
where $\rho = N/(LWH)$ is the mean particle number density. Substituting $j$ into Fourier's law, the thermal conductivity is given by~\cite{PhysRevE.103.L050102}:
\begin{equation}\label{Eq5}
\kappa = L \, \sqrt{\frac{8\rho^{2} k_{\mathrm{B}}^{3}}{\pi m}} \, \bigg/ \left( \frac{1}{\sqrt{T_0}} + \frac{1}{\sqrt{T_L}} \right).
\end{equation}
In Fig.~\ref{fig1}(a) and its inset, the analytical results for $j$ and $\kappa$ (black solid lines) agree perfectly with our simulation data (black circles). As shown, $j$ is constant and $\kappa$ diverges linearly with $L$, consistent with our numerical simulations. \par

We now turn to interacting systems with collisions to investigate how heat conduction depends on interaction strength. The interaction strength is tuned via the collision time interval $\tau$ between successive MPC steps. In our simulations, we fix $H$ and set $W = L$, and study how the thermal conductivity varies as the system size $L$ increases. As $H$ increases from 1 to $L$, a dimensional crossover from q-2D to 3D is realized. As shown in Fig.~\ref{fig1}(a) and its inset, for weak interactions ($\tau = 10$), across the transition from q-2D to 3D, the thermal current scales as $j \sim L^{-1}$ for sufficiently large $L$. Consequently, the thermal conductivity $\kappa$ becomes independent of $L$, consistent with Fourier's law. This provides strong evidence that, in the nearly integrable regime, the system exhibits normal heat conduction. We note that this behavior---observed throughout the q-2D--3D crossover---is governed by the kinetic regime, similar to prior reports in 1D systems~\cite{Chen2014,zhao2018,Fu2026}.\par

\begin{figure}[t]
  \centering
  \includegraphics[width=1\columnwidth]{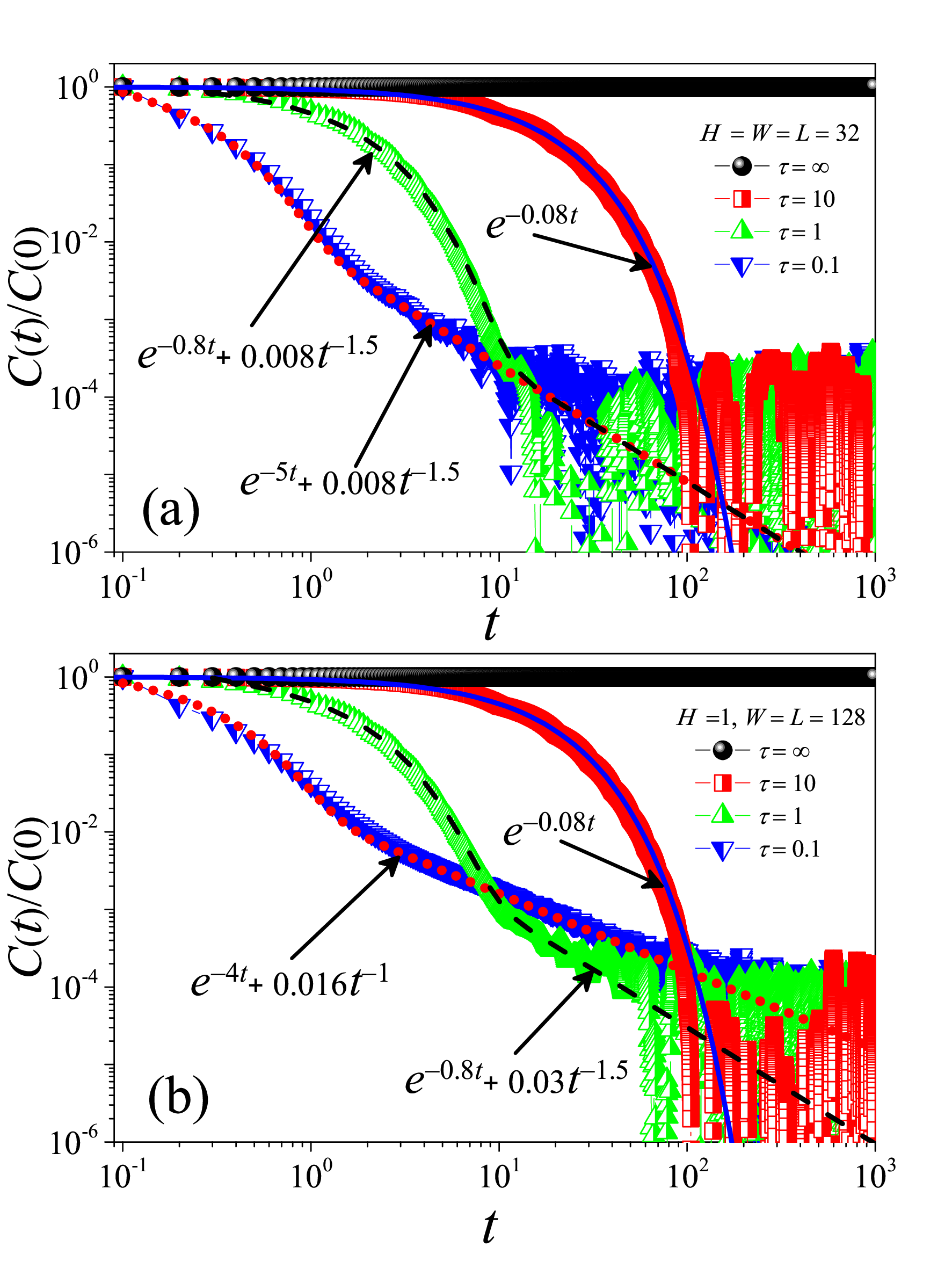}
  \caption{Heat current correlation functions $C(t)$ for the 3D system (a) ($H = W = L = 32$) and the q-2D system (b) ($H = 1$, $W = L = 128$), computed for different values of $\tau$. In both panels, the blue solid, black dashed, and red dotted curves represent the best fits to the data obtained for $\tau = 10$, $\tau = 1$, and $\tau = 0.1$, respectively. Simulation parameters: $m = k_\mathrm{B} = T = a = 1$, $\theta = \pi/2$, and $\rho = 6$.
  }\label{fig2}
\end{figure}

However, in Fig.~\ref{fig1}(b), for stronger interactions ($\tau = 0.1$), the scaling $j \sim L^{-1}$ (or size-independent $\kappa$) is observed only in the 3D system. As the system approaches the q-2D case, the thermal current ultimately follows $j \sim (\ln L)/L$, and the thermal conductivity exhibits $\kappa \sim \ln L$---a scaling predicted for 2D momentum-conserving systems~\cite{Basile2006}. This logarithmic divergence reflects anomalous heat conduction governed by the hydrodynamic regime, a hallmark of nonequilibrium transport in low dimensions~\cite{lepri2016}. These results confirm normal heat conduction in the 3D fluid system and indicate that reducing the ratio of the transverse system size to its length drives a dimensional crossover from 3D to 2D behavior in heat transport.\par

Collectively, the simulation results for heat conduction reveal a crossover from ballistic to kinetic and ultimately to hydrodynamic transport, governed by the MPC collision time scale $\tau$.\par

\emph{Equilibrium results}---To validate the results from non-equilibrium simulations, we compare them with those obtained from linear-response theory in equilibrium. According to the Green--Kubo formula, which relates transport coefficients to the heat current time-correlation function $C(t) \equiv \langle J(0)\,J(t) \rangle$, the thermal conductivity is given by~\cite{1991Kubo,LEPRI20031,Dhar01092008}
\begin{equation}\label{EqGKKL}
\kappa_{\mathrm{GK}} = \frac{\rho}{k_{\mathrm{B}} T^2} \lim_{\tau_{\mathrm{tr}} \to \infty} \lim_{N \to \infty} \frac{1}{N} \int_0^{\tau_{\mathrm{tr}}} C(t) \, dt,
\end{equation}
where the total heat current along the $x$-direction is defined as $J = \frac{1}{2} \sum_{i=1}^{N} m_i v_i^2 v_{x,i}$. In our simulations, we consider an isolated fluid with periodic boundary conditions in all three directions. The system is initialized with zero total momentum and a total energy corresponding to temperature $T$. After equilibration, $C(t)$ is computed from time averages. The upper integration limit is set to $\tau_{\mathrm{tr}} = L / v_s$, where $v_s$ denotes the sound speed. Previous studies have shown that, for sufficiently large systems, the equilibrium thermal conductivity $\kappa_{\mathrm{GK}}$ agrees quantitatively with the non-equilibrium value $\kappa$, provided that $v_s$ is accurately extracted from the spatiotemporal correlation of local heat currents~\cite{Casati2003,Zhao2006,PhysRevE.89.022111}.\par

The correlation functions $C(t)$ for the 3D system ($H = W = L = 32$) and the q-2D system ($H = 1$, $W = L = 128$) are shown in Figs.~\ref{fig2}(a) and~\ref{fig2}(b), respectively, for various values of $\tau$. In the noninteracting case ($\tau = \infty$), $C(t)$ remains constant in time, confirming ballistic transport, as particles traverse the system without collisions, retaining their initial velocities.\par

For the 3D system with weak interactions ($\tau = 10$), Fig.~\ref{fig2}(a) shows that $C(t)$ decays exponentially at early times and, for $t > 10^2$, begins to oscillate about zero; negative values are not shown, as the logarithmic scale requires $C(t) > 0$. This behavior characterizes normal heat conduction in the kinetic regime, consistent with observations in 1D systems~\cite{Chen2014,zhao2018,Fu2026}. In contrast, for stronger interactions ($\tau = 0.1$), $C(t)$ exhibits an initial exponential decay followed by a power-law tail, $C(t) \sim t^{\gamma}$, with $\gamma = -1.5$, in excellent agreement with the hydrodynamic prediction for 3D systems~\cite{lepri2016,Basile2006}. These results thus demonstrate that, as $\tau$ is decreased from 10 to 0.1, the mechanism underlying normal heat conduction in the 3D system---observed in Fig.~\ref{fig1}---undergoes a transition from kinetic to hydrodynamic dominance.\par

A similar phenomenon is observed in the q-2D system (Fig.~\ref{fig2}(b)): as $\tau$ is reduced from 10 to 1, $C(t)$ evolves from exponential to power-law decay, $C(t) \sim t^{\gamma}$ with $\gamma = -1.5$, indicating a transition from kinetic to hydrodynamic dominance in normal heat conduction. Remarkably, upon further decreasing $\tau$ to 0.1, the asymptotic decay exponent becomes $\gamma = -1$, fully compatible with the theoretical prediction for the 2D systems~\cite{lepri2016,Basile2006}. Integrating $C(t) \sim t^{-1}$ in Eq.~(\ref{EqGKKL}) yields a superdiffusive thermal conductivity $\kappa_{\mathrm{GK}} \sim \ln L$, in agreement with the non-equilibrium simulation data shown in the inset of Fig.~\ref{fig1}(b). This provides compelling evidence for a dimensional crossover in heat transport---from three to two dimensions---as the system thickness is reduced, thereby reducing the aspect ratio $H/L$.\par

The equilibrium results presented above indicate that, in the momentum-conserving MPC fluids, the thermal transport mechanism---governed by the MPC collision time scale $\tau$---consistently undergoes a transition from ballistic to kinetic and ultimately to hydrodynamic, irrespective of the system dimensionality (q-2D or 3D). These results fully corroborate the behavior observed in non-equilibrium systems.\par

\emph{Summary and discussion}---In summary, we have investigated heat conduction in a momentum-conserving MPC fluid across the dimensional crossover from q-2D to 3D. By tuning the interaction strength via the MPC collision time interval $\tau$, we identify three distinct transport regimes: \\
(i) ballistic transport, where $\kappa \sim L$ and $C(t)$ is constant; \\
(ii) kinetic transport, characterized by a size-independent $\kappa$ and exponential decay of $C(t)$; \\
(iii) hydrodynamic transport, where the heat conduction behavior bifurcates with dimensionality: the q-2D system exhibits $\kappa \sim \ln L$ and $C(t) \sim t^{-1}$, while the 3D system shows a size-independent $\kappa$ and $C(t) \sim t^{-3/2}$.

Our results for q-2D to 3D systems, together with those previously reported in 1D systems~\cite{Zhong2012,Zhong2013,Chen2016,Chen2014,zhao2018,Fu2026}, collectively demonstrate that normal heat conduction dominated by the kinetic regime is a universal phenomenon. Although prior evidence~\cite{Fu2026,Wang2013,Das2014,Miron2019,Lepri2020,lepri2026} suggested that this phenomenon arises from finite-size effect, these results provide a theoretical foundation for understanding the normal heat conduction observed in finite-size real systems. Furthermore, our findings---observed in the hydrodynamic regime---quantitatively validate the scaling predictions~\cite{Lepri1998,Narayan2002,Basile2006} for heat transport and reveal a clear dimensional crossover from 2D-like anomalous transport to 3D Fourier behavior, highlighting dimensionality as a critical control parameter for size-dependent thermal transport. These results significantly advance the understanding of heat conduction and directly address the question raised in the Introduction.

\begin{acknowledgments}
This work is supported by the National Natural Science Foundation of China (Grant No.12475034 ) and the Natural Science Foundation of Fujian Province (Grant No.2023J05100).
\end{acknowledgments}

\bibliography{HCRefs}

\end{document}